\documentclass[journal, comsoc]{IEEEtran}

\usepackage[T1]{fontenc}
\usepackage[english]{babel}
\usepackage[ruled, vlined]{algorithm2e}
\usepackage{algpseudocode}
\usepackage{amssymb}
\usepackage{array}
\usepackage{arydshln}
\usepackage{blkarray, bigstrut}
\usepackage{bm}
\usepackage{booktabs}
\usepackage{braket}
\usepackage{cite}
\usepackage{cuted}
\usepackage{enumitem}
\usepackage{graphicx}
\usepackage[mathscr]{euscript}
\usepackage[switch]{lineno}
\usepackage{mathtools}
\usepackage{multirow}
\usepackage{multicol}
\usepackage{mathtools}
\usepackage{xfrac}
\usepackage{framed} 
\usepackage[framed]{ntheorem}
\usepackage{pgfplots}
\usepackage{qtree}
\usepackage{adjustbox}
\usepackage{rotating}
\pgfplotsset{compat = 1.12}
\usepackage{stfloats}
\usepackage{subcaption}
\usepackage{url}
\usepackage{tikz}
\usepackage{tkz-berge}
\usepackage[framemethod=TikZ]{mdframed}
\usetikzlibrary{patterns}
\usetikzlibrary{spy}
\usetikzlibrary{shapes, backgrounds,calc, snakes}
\usetikzlibrary{decorations.pathreplacing}
\usetikzlibrary{matrix}
\usepackage{fancybox}
\usepackage[normalem]{ulem}
\usepgfplotslibrary{fillbetween}
\usepackage[colorinlistoftodos,prependcaption,textsize=scriptsize]{todonotes}

\makeatletter
\def\newmaketag{%
	\def\maketag@@@##1{\hbox{\m@th\normalfont\normalsize##1}}%
}
\makeatother

\usepackage{todonotes}

\newcommand{\ally}{\textcolor{black}}

\newcommand{\fref}[1]{Fig.~\ref{#1}}

\newcommand{\tref}[1]{Table~\ref{#1}}

\begin{document}
\title{6G Massive Radio Access Networks: Key Issues, Technologies, and Future Challenges}

\author{Ying Loong Lee,~\IEEEmembership{Member,~IEEE},
       Donghong Qin,~\IEEEmembership{Member,~IEEE}, Li-Chun Wang,~\IEEEmembership{Fellow,~IEEE}, Gek Hong (Allyson) Sim,~\IEEEmembership{Member,~IEEE}
\thanks{Y. L. Lee is with the Universiti Tunku Abdul Rahman, Cheras 43000, Kajang, Selangor, Malaysia (e-mail: leeyingl@utar.edu.my).}
\thanks{D. Qin is with the Guangxi University for Nationalities, 530004 Nanning, China (e-mail: donghong\_qin@163.com).}
\thanks{L.-C. Wang is with the National Chiao Tung University, Hsinchu 300, Taiwan. (e-mail: lichun@g2.nctu.edu.tw).}
\thanks{G. H. Sim is with the Technische Universität Darmstadt, 64289 Darmstadt, Germany (e-mail: sallyson@seemoo.tu-darmstadt.de).}
\thanks{Manuscript received xxx, 2019.}
\thanks{}\thanks{}
}

\markboth{Journal of \LaTeX\ Class Files,~Vol.~XX, No.~XX, XX~XX}%
{Shell \MakeLowercase{\textit{et al.}}: Bare Demo of IEEEtran.cls for IEEE Communications Society Journals}

\maketitle

\begin{abstract}
Driven by the emerging use cases in massive access future networks, there is a need for technological advancements and evolutions for wireless communications beyond \ally{the} fifth-generation (5G) networks. In particular, we envisage the upcoming sixth-generation (6G) networks to consist of numerous devices demanding extremely high-performance interconnections even under strenuous scenarios such as diverse mobility, extreme density, and dynamic environment. To cater \ally{for} such \ally{a} demand, investigation on flexible and sustainable radio access network (RAN) techniques capable of supporting highly diverse requirements and massive connectivity is of utmost importance. 
To this end, this paper first outlines the key driving applications for 6G, including smart city and factory, which trigger the transformation of existing RAN techniques. We then examine and provide in-depth \ally{discussions} on several critical performance requirements (i.e., the level of flexibility, the support for massive interconnectivity, and energy efficiency), issues, enabling technologies, and challenges in designing 6G massive RANs. We conclude the article by providing several artificial-intelligence-based approaches to overcome future challenges.
\end{abstract}


\begin{IEEEkeywords}
6G, flexibility, massive access, energy efficiency, massive interconnectivity
\end{IEEEkeywords}

\IEEEpeerreviewmaketitle

\section{Introduction}

\IEEEPARstart{T}{he} global mobile \ally{Internet and telecommunication} traffic is expected to reach 77 exabytes per month by 2022 -- a seven-fold increase over 2017,
as predicted by Cisco. Driven by the rapid growth in mobile traffic, the fifth-generation (5G) wireless communications have been extensively researched in meeting the key performance requirements for enhanced mobile broadband (eMBB), ultra-reliable and low-latency communications (URLLC), and massive machine-type communications (mMTC)~\cite{ofdm_iot}. As a result, \ally{the deployment of 5G networks will be widespread by 2020.} 
However, as mobile traffic continues to \ally{grow}, 5G will eventually encounter technical limitations in supporting \emph{massive interconnectivity with highly diverse service requirements}. Furthermore, a number of emerging use cases and dynamic future scenarios foster the need for a new paradigm \ally{for the} next-generation wireless communications -- sixth-generation (6G). 

Recently, both academia and industry have begun to \ally{investigate} the visions, requirements, scenarios, key technologies, and system architectures for 6G networks~\cite{6g_wireless, 6g_tech, doma, 6g_vision, flex, ofdm_iot, scal_cloud, netslic_survey}. In fact, the international standardization organization, International Telecommunication Union and Telecommunication Standardization Sector~(ITU-T), has formed a study group, namely Focus Group Technologies for Network 2030 (FG NET-2030), to identify and investigate issues, challenges, scenarios, and fundamental technologies required for 5G and beyond. Clearly, the aforementioned initiatives indicate a dire need to begin intensive research and development for 6G wireless communications. This measure is critical in preparation for the unprecedented demands for ubiquitous connectivity in ultra-dense networks.

Although the existing works on 6G have articulated the visions, requirements, and enabling technologies for 6G, several critical aspects on \emph{6G radio access networks (RANs)} have not been discussed intensively, \ally{including} \emph{flexibility, massive interconnectivity, and energy efficiency}. These aspects play a vital role in guaranteeing the scalability and sustainability of future ultra-dense networks which support a multitude of interconnectivity that fulfill the diverse service requirements.
To this end, in this article, we first highlight key driving applications for 6G and motivate the need to look into the above-mentioned aspects. Then, we provide in-depth discussions on each aspect, identify the corresponding issues, and investigate the potential technologies to address them. Lastly, we reveal several future challenges and conclude the article with several future directions for highly flexible and energy-efficient 6G massive RANs.

\IEEEpubid{Note: This work has been submitted to the IEEE for possible publication. Copyright may be transferred without notice, after which this version may no longer be accessible.
}
\IEEEpubidadjcol
\section{Key Driving Applications for 6G-RAN}
Several emerging applications and future use cases have become the key driving factors for 6G. In this section, we outline the critical use cases for 6G networks and the need to rethink new RAN designs for each of them. These use cases include multimedia and entertainment, smart city, smart industry, and beyond-terrestrial communications.

\begin{table*}[bh!]
\caption{Summary on 6G key driving factors.}
\centering
\begin{tabular}{lll}
\hline\hline
Key Driving Factors  & Example Applications / Scenarios & Requirements \\
\hline\hline
Multimedia and Entertainment & \begin{tabular}[c]{@{}l@{}}AR, XR, VR, cloud gaming, super HD audio-video streaming\end{tabular}                                                                                          & \begin{tabular}[c]{@{}l@{}}$\bullet$ Super-high throughput\\$\bullet$ Ultra-high bandwidth\end{tabular}                                                                                                             \\
\hline
Smart City & \begin{tabular}[c]{@{}l@{}}Connected vehicles, autonomous driving, public surveillance, smart healthcare, \\ intelligent transportation, smart tourism, UAVs, drones\end{tabular} & \begin{tabular}[c]{@{}l@{}}$\bullet$ Ultra-high bandwidth\\$\bullet$ Ultra-low latency\\$\bullet$ Ultra-high reliability\\$\bullet$ Extremely high data density\\$\bullet$ Super-high intelligence\\$\bullet$ Ultra-high energy efficiency\end{tabular} \\
\hline
Smart Industry & \begin{tabular}[c]{@{}l@{}}Autonomous manufacturing, smart production, remote diagnostics,\\ remote robotic control, connected goods\end{tabular} & \begin{tabular}[c]{@{}l@{}}$\bullet$ Ultra-low latency\\$\bullet$ Ultra-high reliability\\$\bullet$ Extremely high data density\\ $\bullet$ Very high intelligence\\$\bullet$ Very high energy-efficiency\end{tabular}                          \\\hline
Beyond-Terrestrial Communications & \begin{tabular}[c]{@{}l@{}}Airplanes, satellites, ships communications, undersea sightseeing,\\ submarines\end{tabular}                            & \begin{tabular}[c]{@{}l@{}}$\bullet$ Ultra-high bandwidth\\$\bullet$ Ultra-long distance\end{tabular}  \\
\hline\hline
\label{tab1}
\end{tabular}
\end{table*}

\noindent{\textbf{Multimedia and Entertainment. }}
Virtual reality (VR), augmented reality (AR), extended reality (XR), cloud gaming, and ultra-high-definition (HD) audio-video streaming applications have received a popularity boost in the multimedia and entertainment industry. Some of these applications will incorporate high-precision tactile and haptic features (e.g., touch and smell) to provide a close-to-reality experience to users \cite{6g_wireless}. Such applications \ally{require} extremely high throughput requirements in the order of Gbps to Tbps. Besides, large amounts of communications, sensing, data processing,  and control \ally{are subject to stringent delay to support strict real-time services.}

\noindent{\textbf{Smart City. }}
The advancement in the Internet-of-Things (IoT) technologies has realized several applications for smart cities, such as intelligent transportation, autonomous driving, connected vehicles, smart healthcare, public surveillance, and smart tourism. As these applications involve large numbers of wirelessly connected sensors, cameras, caches, computing elements, and controllers, massive amounts of data and transmissions will take place in every corner of the cities which are equipped with those devices. As such, ultra-broadband coverage and ultra-high bandwidth will be needed to support the full connectivity of the \ally{IoT} devices everywhere in the cities and the transmission of massive data. Some applications such as autonomous driving, unmanned aerial vehicles (UAVs), and smart healthcare will require stringent latency and ultra-high reliability communications for precise monitoring, control, and coordination, especially in emergency situations. In particular, the use of UAV-based base stations (BSs) would require the RAN that can adapt to the change in the network for fulfilling the target performance requirement. Furthermore, the management of network resources, functions, and controls for the massive interconnectivity in smart cities will need to be \ally{automated}. Therefore, it is vital that 6G RANs operate intelligently where RAN nodes can autonomously, collectively, and cooperatively coordinate among each other to manage the massive interconnectivity, while taking \ally{account of the requirements} for bandwidth, latency, and reliability. Given the massive amounts of data processing and computations in such intelligent systems, smart city applications are expected to consume an excessive amount of energy at both the device and network levels. To this end, a system-wide ultra-high energy-efficient RAN will be the key performance requirement for 6G.

\noindent{\textbf{Smart Industry. }}
\ally{Fostered} by the advances in the industrial IoT technologies and cyber-physical systems, the industrial sector has begun to move toward the automated industry (i.e., Industry 4.0). Industrial automation encompasses a wide range of applications such as smart production, autonomous manufacturing, connected machines, and intelligent remote robotic control. With the rapid advancement of wireless technologies, wireless remote execution of these applications, which facilitate the automated industrial processes, becomes feasible. However, the challenge lies in efficiently coordinating the interconnections among the massive number of devices in smart factories (e.g., sensors, cameras, machines, robots, and even goods) in order to facilitate highly reliable and ultra-low latency communications. Specifically, applications involving high-precision controls and instant responses (e.g., robotics, augmented reality-based remote control, and maintenance for time-critical factory processes) require extremely low-latency communications and high-performance computing to respectively deliver and process the massive amount of data that will be generated in future smart factories. 
Given the high energy consumption in processing these massive data, existing energy-efficient techniques will require significant improvement.

\textbf{Beyond-Terrestrial Communications}:
Next-generation mobile networks are envisioned to cover various beyond-terrestrial domains such as seas, underwater, aerial, and space \cite{6g_wireless, 6g_tech}. Relevant use cases include communications on the airplanes, satellites, ships, and undersea submarines. Such communications will require support for ultra-long-distance transmissions.


\tref{tab1} summarizes the key driving factors for 6G, its example applications or scenarios, and their requirements.

\section{Key Requirements for 6G RAN}
\ally{In the future, 6G is envisioned} to support massive interconnectivity, offer advanced computation capability, provide high data rates even at high mobility and long distances, and accommodate the exchange of immense amounts of data. As such, RAN flexibility and the capability to support massive interconnectivity are critical in the development of key technologies for 6G RAN. In addition, energy efficiency for 6G RAN requires more emphasis than that \ally{for} 5G RAN because the future use cases will involve very high energy consumption. To that end, we dedicate this section to elaborates on several fundamental performance requirements for 6G RAN, namely, \emph{flexibility, massive interconnectivity, and energy efficiency}.

\subsection{Flexibility}
6G needs to be fully flexible, that is, the ability to adaptively customize the RAN configurations, functions, and resources to accommodate the various requirements of the diverse applications and services as well as the spatially varying user densities and traffic demands. In what follows, we discuss three important features that facilitate RAN flexibility: flexible frame structure, scalable cell size, and flexible placement of logical RAN functions and scaling of RAN resources.


5G has adopted a flexible transmission frame structure, where the orthogonal division frequency multiplexing (OFDM)-based subframes are parameterized based on multiple OFDM numerologies \cite{5Gspec}.
By parameterizing the OFDM-based subframes, the subcarrier spacing of resource blocks (RBs)\footnote{Each RB consists of 12 OFDM subcarriers.} can be scaled according to $15\times 2^n$ kHz, where $n$ is the integer-valued numerology parameter. The subcarrier spacing can be chosen according to performance (e.g., latency, reliability, throughput, etc.)  requirements of the specific application. For instance, wider subcarrier spacings are suitable for low-latency and high-reliability critical applications such as autonomous driving and UAVs, whereas narrower subcarrier spacing is suitable for low data rate or machine-type communications such as narrowband IoT applications. In addition, a wider subcarrier spacing is essential for applications operating at higher frequency bands, such as millimeter wave (mmWave), to alleviate the Doppler spreads for high-mobility scenarios.

Clearly, the scalable OFDM numerology proposed for 5G will be the underlying foundation for designing flexible 6G RAN\cite{flex}. However, it does not address spatial problems such as different user densities and heterogeneous traffic demands in different geographical areas. These issues have nonetheless been addressed in cellular networks, where system-level flexibility is achieved through the ultra-dense deployment of small-cells. 
Precisely, in areas with high user densities, the coverage of the nearby small cells can be expanded by increasing their transmit power. On a finer scale, the transmit power on the RBs allocated for a user can be adapted to guarantee its quality of service (QoS). The accessible bandwidth of each small cell can be adjusted to meet the variation in traffic demands from the users in proximity. 
Such system-level flexibility potentially complements the scalable OFDM numerology to provide an even higher level of flexibility for radio access for  heterogeneous traffic demands and requirements in different areas, as depicted in  \fref{fig:flexi_ran}. 


\ally{Another technology contributing to flexible RAN is cloud and edge computing. Currently, cloud and edge computing allows functional splits between RAN functions so that some of the RAN functions are centralized in the cloud while others are located at the radio frontend or edge server. By leveraging network virtualization, logical RAN functions (e.g., scheduling, baseband processing)  can be flexibly virtualized, scaled, and placed at the cloud or at the edge based on service types, QoS requirements, or transport network requirements \cite{scal_cloud}. For instance, more storage and computing resources can be allocated for logical RAN functions in the control plane (CP) than in the user plane (UP) for machine-type applications \ally{because} these applications often generate small data packets with a multitude of connections \cite{scal_cloud}. Moreover, the computational capacity of baseband processors can be scaled according to the QoS requirements, such as minimum data-rate requirements \cite{yllee}. RAN functions that handle strict real-time (RT) requirements such as scheduling can be placed close to the radio interface (e.g., radio frontend and edge server) to deal with low-latency applications. Besides, for more latency- and reliability-critical applications that heavily involve a number of CP functions, these functions may be moved to the BSs' radio frontend or edge server. Top-right of \fref{fig:flexi_ran} illustrates the placement of the logical RAN functions between the cloud and the radio frontend/edge server. }

\subsubsection{Key Issue}
Current research on flexible RANs focus on three main directions: scalable OFDM numerologies, cell size scaling, and flexible scaling and placement of RAN functions. Since 6G RAN is expected to be fully flexible, the combination of these directions is critical. For each typical use case, 6G should be able to flexibly customize the appropriate OFDM numerology, transmit power level, computing resources, and the placement of logical RAN functions. Moreover, flexible radio access needs to consider a multitude of technical factor, including  service types, QoS requirements, channel conditions, carrier frequencies, user locations, availability of cloud computing resources, transmit power budget, RB availability, statistical multiplexing gain of the central cloud, and transport network requirements~\cite{scal_cloud}. For use cases with flying UAV-based BSs, more control parameters (e.g., UAVs' positions, antenna beam angles, etc.) have to be considered to adapt to the network dynamics and meet the performance requirements.

\subsubsection{Enabling Technology}
\begin{figure*}[t!]
\centering
\begin{subfigure}{\textwidth}
	\centering
    \includegraphics[width=\columnwidth]{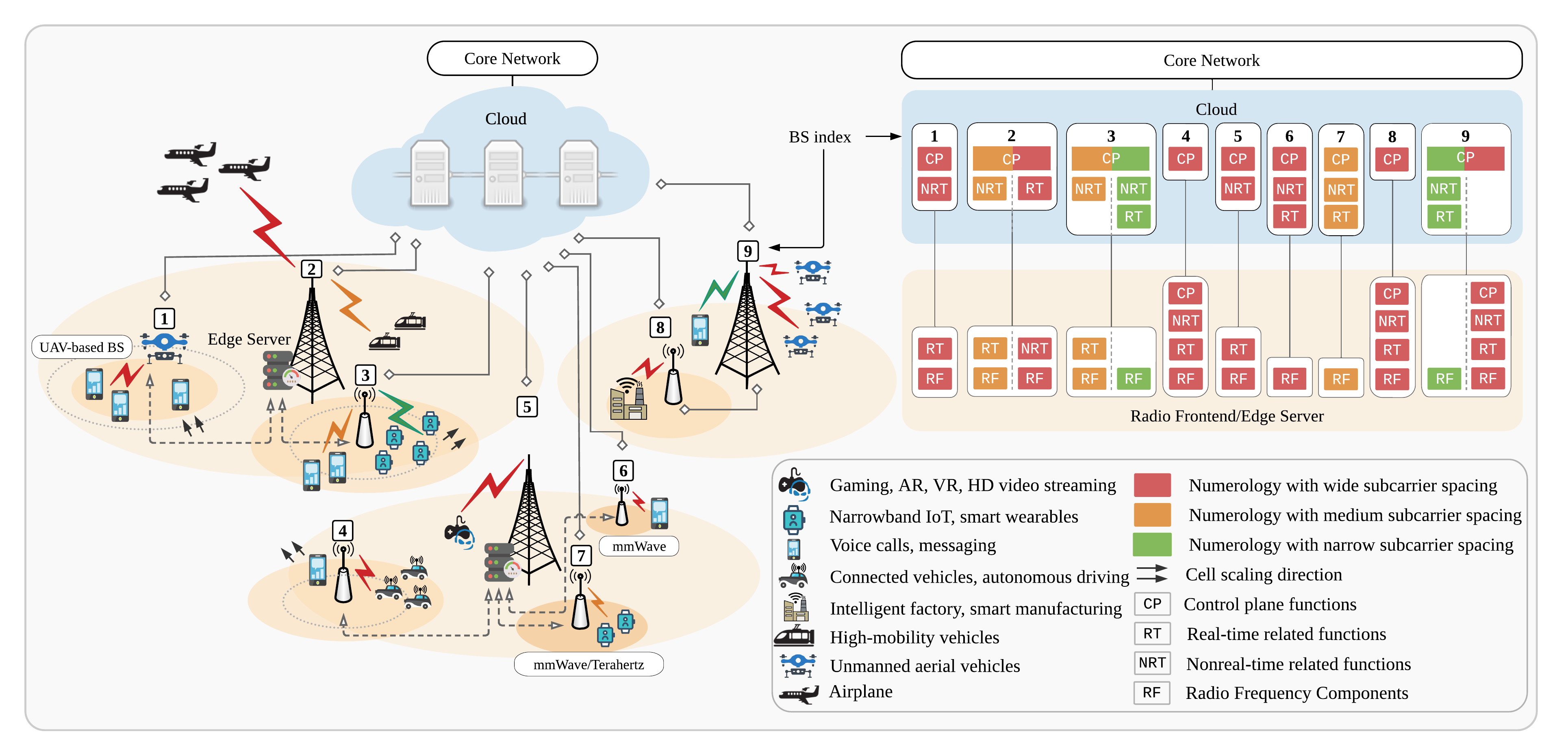}
	\caption{(Left) Flexible RAN architecture with scalable OFDM numerologies and cell sizes. (Top-right) The corresponding flexible placement of logical RAN functions at the cloud and radio frontend}
	\label{fig:flexi_ran}
\end{subfigure}
\hspace{2mm}
\begin{subfigure}{\textwidth}
	\centering
    \includegraphics[width=0.75\columnwidth]{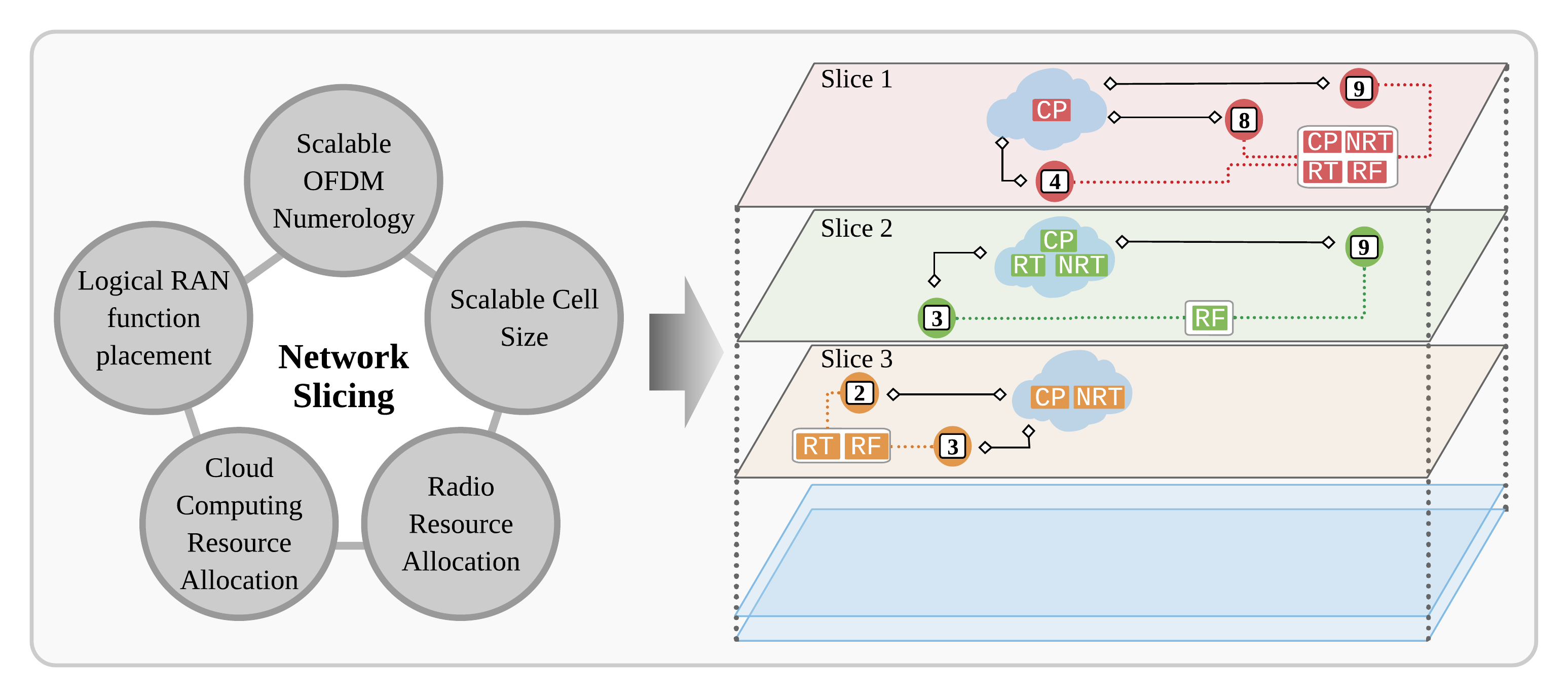}
    \vspace{-2mm}
	\caption{RAN slices with specific radio access settings}
	\label{fig:ran_slice}
\end{subfigure}
\caption{Our vision for 6G flexible RAN architecture, placement of logical RAN functions, and network slicing.}
\label{fig:6g_ran}
\vspace{-7mm}
\end{figure*}

Network slicing \cite{netslic_survey} is a promising technique that fully leverages flexibility. In the context of RAN, the concept of network slicing defines the process of virtualizing multiple logical networks containing all essential (virtual) RAN functions and resources. Each of these \emph{network slices} can be customized to a  specific use case or service. Having said that, 6G network slices can be instantiated with specific settings of OFDM numerology, placement of RAN functions, transmit power, and bandwidth of the involved cells, based on the aforementioned technical factors for flexible radio access. \fref{fig:ran_slice} shows several network slices instantiated for different types of services. For example, slice 1 is instantiated for services related to latency-critical applications (e.g., UAVs and autonomous driving). The CP and non-RT related RAN functions are placed at the cloud, while the RT-related RAN functions are placed at the \ally{radio frontend/edge server}, where an OFDM numerology with wide subcarrier spacings is selected. Cell size scaling can be performed by the RF component at the \ally{radio frontend/edge server}. Slice 2 deals with low-rate communications. Thus, most of the RAN functions are located at the cloud. Slice 3 is instantiated for applications with medium QoS requirements and for cells with a small coverage. Multiple network slices can be instantiated and orchestrated on the physical network infrastructure in the network by leveraging the network function virtualization (NFV) and software-defined networking techniques, which have recently been regarded as enablers for network slicing \cite{netslic_survey}. 

\subsection{Support for Massive Interconnectivity}
Emerging smart city and industrial IoT applications involve vast amounts of communications, data exchange, sensing, storage, controls, and computing in supporting the upcoming massive interconnectivity in dense 6G networks. Indeed, supporting such connectivity requires sufficient capacity and efficient access techniques. 

Network densification is a viable way to support massive interconnectivity. By densely deploying low-power BSs, network capacity increases and more users/devices can be supported. Capacity can be further enhanced by exploiting the spectrum beyond sub-6GHz bands such as mmWave, which provide a substantial amount of bandwidth to support data-intensive use cases such as: \emph{(i)} the delivery of large-size sensory data in vehicular communications, \emph{(ii)} very-low latency and data-rate intensive sensory data delivery in XR as well as UAV applications in smart city, and \emph{(iii)} connected robotics in Industry 4.0.

\subsubsection{Key Issue}
Network densification alone cannot keep up with the rapid growth in the number of wireless devices. In addition, inter-cell interference can severely limit the achievable capacity when BSs are in proximity. Although exploiting the mmWave or Terahertz band potentially provides extremely-high data rates and capacity, it is only suitable for short-range communications due to severe path loss issues. In addition, the need for appropriate beamforming and beam-alignment coordination among the interconnecting devices (to avoid interference) complicate the access procedure. Thus, it is hard to guarantee reliable communications. This is particularly problematic as enabling future super-smart city and industry will require ultra-high reliable communications providing data rates of up to 1 Tbps under dynamic and dense scenarios.

\begin{figure*}[!t]
\centering
\includegraphics[width=0.95\textwidth]{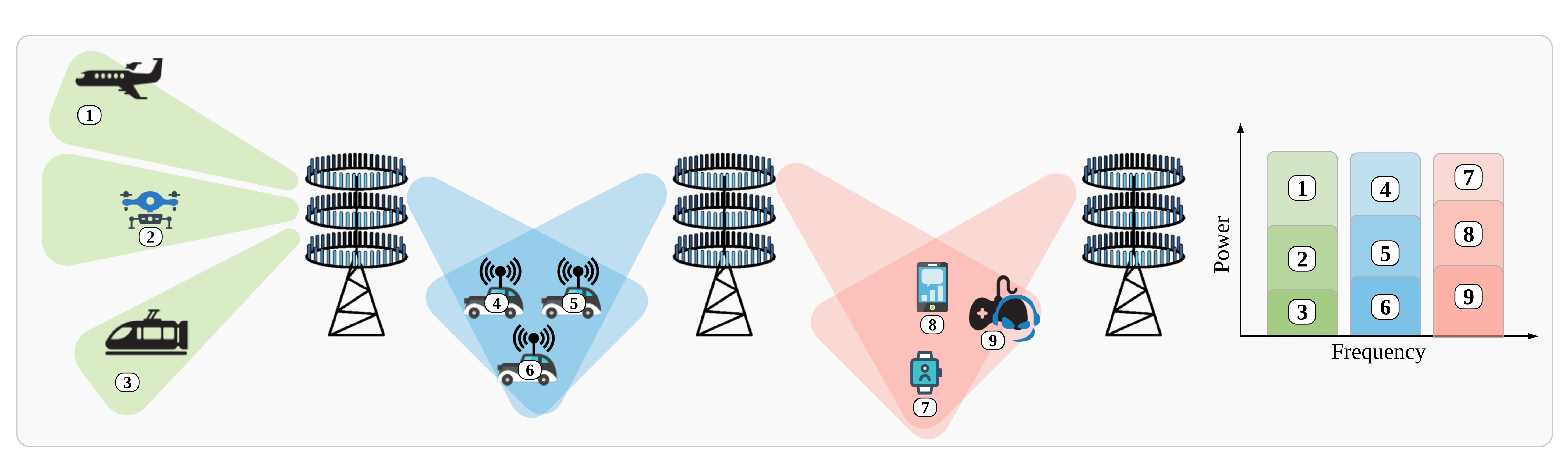}
\caption{BS cooperation with mMIMO-NOMA for massive connectivity with diverse applications' requirements.}
\label{fig2}
\end{figure*}

\subsubsection{Enabling Technologies}
Recently, non-orthogonal multiple access (NOMA) has shown great potential in accommodating massive connectivity. By allocating a different transmit power level to each signal, the scheme simultaneously transmits multiple signals in one frequency subband. The signal can be retrieved by applying successive interference cancellation (SIC), which first decodes the signal with the highest power by treating the other signals as interference, followed by the signal with the next highest power, and so on. Unlike orthogonal multiple access (OMA) schemes, NOMA can accommodate a much larger maximum number of connections. 

Current research on NOMA shows its effectiveness in sustaining massive machine-type and IoT communications. However, NOMA lacks the flexibility to support the massive connectivity with highly diverse service requirements~\cite{flex}. Such scenarios can be found under various industrial automation use cases. For example, delay-critical factory process optimization poses high reliability, low latency and small indoor coverage requirements with demands of heavy data transfer, whereas remote quality inspections and diagnostics require high-reliability and wide coverage conditions. A basic approach to improve NOMA for QoS provisioning is to dynamically allocate a suitable power level to each of the multiple signals to be transmitted in such a way that their minimum data rates can be met. However, the power allocation is limited by the power budget of the serving BS. The study in \cite{doma} proposes a massive in-band NOMA, in which multiple BSs can cooperatively transmit to multiple users using the same frequency subband. This approach is feasible because the next-generation network will continue to be densified, and it can cooperate with several BSs to increase capacity. With careful selection of multiple BSs and power allocation among users/devices, massive amounts of connections with diverse requirements can be supported. This technology can also be supplemented by the flexible radio access technology discussed earlier \cite{flex}.

Massive MIMO (mMIMO) is another candidate technology to enable massive access. By equipping the BS with a large number of independently-controlled antennas, multiple data signals can be multiplexed in the spatial domain by transmitting signals on directional antenna beams in the same frequency subband. On top of that, by applying advanced rate splitting and hybrid-precoding techniques in~\cite{abantoleon:2019:hybrid} to manipulate the beamforming, high spectral efficiency, reliable, and low power consumption solution can be achieved in high-density networks. 
This is beneficial for industrial IoT applications where many devices will be concentrated in the factory, which makes mMIMO beamforming very suitable. In \cite{miot}, it has been shown that mMIMO with 400 antennas can cater to more than 8,000 industrial IoT devices with low delay, moderate rate requirements - a compelling case where the number of antennas is smaller than the number of devices. Furthermore, mMIMO can interwork with massive in-band NOMA to further boost the network capacity and mitigate interference between users in the spatial and power domains. Fig. \ref{fig2} illustrates a network scenario, where mMIMO and massive in-band NOMA are jointly enabled to support massive connectivity with various application requirements.  


\subsection{Energy Efficiency}
Smart city and industrial IoT applications involve large amounts of data transmission and processing, which consume high energy levels. As such, 6G needs to be able to save energy while effectively supporting the massive connectivity. A number of studies have been carried out to minimize the energy required to deliver satisfactory network performance by efficiently allocating radio resources, transmit power, and backhaul capacity, as well as cell on/off switching. 

\subsubsection{Key Issue}
The advantage of efficient resource allocation and cell on/off switching diminishes with the rapid increase of network nodes (e.g., BSs, mobile users, IoT devices). Furthermore, this approach typically involves solving a centralized optimization problem, which is often time-consuming and impractical when considering numerous BSs/users/devices. In the case of network densification, the overwhelming inter-cell interference can limit the upper bound of achievable energy efficiency. 

\subsubsection{Enabling Technology}
\begin{figure}[!t]
\centering
\includegraphics[width=0.5\textwidth]{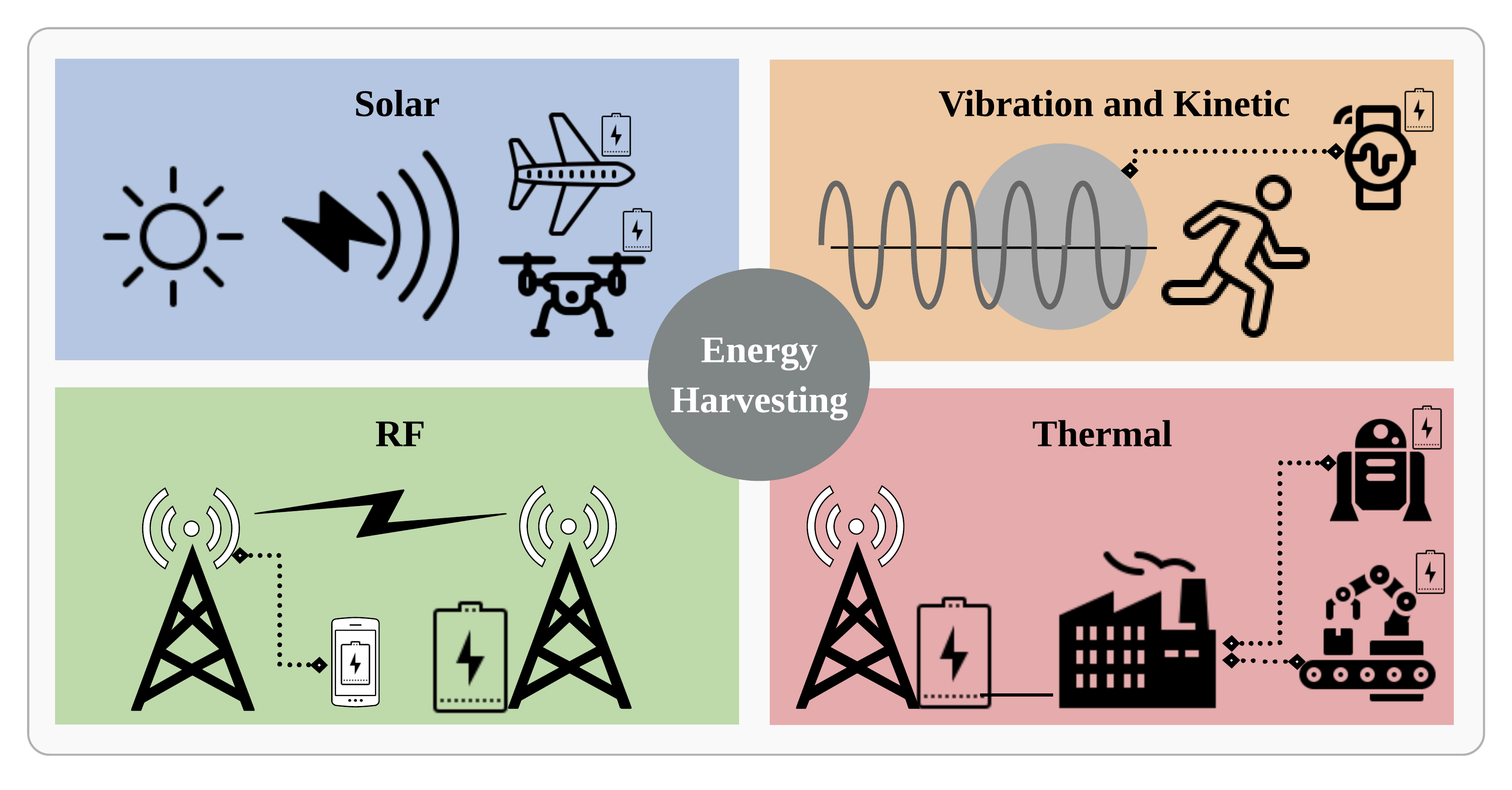}
\caption{Energy harvesting scenarios in future wireless networks.}
\label{fig3}
\end{figure}

Energy harvesting is a promising alternative for energy-efficient massive radio access, where the energy from ambient environments can power network nodes. For example, \ally{as depicted in \fref{fig3}, terrestrial and drone communications can be supported by solar energy. Vibration and kinetic energies can power human body wearables. Industrial IoT can be powered by thermal energy produced from the industrial manufacturing and production process; and ground BSs can be powered by RF energy emitted from other BSs.} Not only does this approach conserve energy on the network infrastructure, but it also prolongs the battery life at the user side. This approach is in particular important since current demands for energy at the user side are gradually increasing due to the proliferation of advanced wireless technologies such as caching and edge computing. 

Current advances in energy harvesting technology allow BSs to realize simultaneous wireless information and power transfer (SWIPT) whereby both information and energy can be extracted from the same received RF signals.
On the other hand, RF interference can be a constant source of energy harvesting. In particular, interference energy harvesting is most useful in ultra-dense networks with massive numbers of network nodes. Besides, artificial noises and jamming signals can also be potential energy sources. Such sources have been considered for wireless energy harvesting \cite{energy}. Energy-efficient and energy-harvesting schemes can be jointly incorporated with massive access technologies for \ally{performance improvement}. For instance, massive MIMO beamforming can help minimize interference for SWIPT-enabled networks.

\section{Future Challenges}
In the previous section, we \ally{elaborated} several promising technologies \ally{that can potentially} be used to tackle specific issues in meeting a particular performance requirement of 6G massive radio access. While promising, there remain challenges that \ally{are} needed to be overcome. 
 
\subsection{Limitations on Flexible Radio Access}
Cell size and carrier frequency can \ally{restrict} the choice of OFDM numerology \cite{ofdm_iot}. On the one hand, the use of numerologies with wider subcarrier spacing is generally more suitable for smaller cell sizes due to their lower delay spreads as compared to that of larger cell sizes. On the other hand, larger cell sizes can also use the numerologies with wider subcarrier spacings, but at the cost of lower throughput. 
It is also important to note that the \ally{cell size} with high-frequency carriers is limited due to path loss and Doppler spread problems in \ally{high-mobility scenarios}.  


\subsection{Resource Allocation for BS Cooperation with mMIMO-NOMA}
BS cooperation with mMIMO-NOMA involves the selection of cooperating BSs, beam-steering of each cooperating BS, and the selected number of data signals in each frequency sub-band. Besides, each cooperating BS needs to jointly allocate power fractions for the selected data signals \ally{to achieve the target} objective (e.g., maximum throughput, the fulfillment of minimum user rate, etc.). This can be modeled as an optimization problem. However, obtaining the optimal solution would be very \ally{difficult} and time-consuming. Therefore, such {an optimization problem} is not suitable to be solved for massive ultra-low latency applications. \ally{Although} solving the problem in a decentralized manner allows to obtain solutions in a shorter time, it is at the cost of lower performance. This trade-off is indeed critical for 6G applications with high-performance demand. Hence, there is an urgent need for a low complexity, efficient, and intelligent solution in coping with continuous network densification in 6G networks.

\subsection{Interference Management for SWIPT}
Interference between BSs and mobile devices is a major challenge for SWIPT, especially for ultra-dense networks because high RF interference may result in difficulty in retrieving the data information. Currently, the  available interference management methods for SWIPT include beamforming, interference alignment, and interference cancellation \cite{energy}. Given the complexity in managing interference in the existing network, it will be even more difficult in 6G networks, where cell coverage \ally{dynamically changes} depending on the application-specific performance requirement, coordination between cooperating BSs, restriction of operating frequency, and many other varying parameters in the network. Furthermore, considering that 6G is envisioned to be a massive access network, it is thus necessary to rethink and redesign the interference management for SWIPT for this network.

\subsection{Throughput-Interference Tradeoff for Energy Harvesting}
Leveraging RF interference for energy harvesting is a challenging problem due to the trade-off relationship between interference and throughput. Specifically, high interference limits the achievable throughput, while low interference limits the amount of RF energy harvested \cite{tradeoff}. Finding the throughput-interference trade-off can be problematic because it is a multi-criteria problem in which multiple (Pareto-)optimal solutions may exist. \ally{This} problem becomes more complicated in ultra-dense massive \ally{RANs}. Distributed suboptimal approaches may be more efficient and feasible for such a problem. 

\subsection{Network Dynamics}
With the introduction of more critical and massive access use cases in the future (e.g., UAVs, drones, automated vehicles, massive IoT), the network dynamics will become significantly more complicated compared to 5G. The network dynamics for the future use cases include network topology, channel condition, traffic demand, spatial user density, and many other factors that have to be considered for supporting massive interconnections with highly diverse QoS requirements. Besides, operating carrier frequencies, especially those in the mmWave and Terahertz, have issues with high-mobility and long-distance transmissions. Hence, careful consideration of the carrier frequency aspect is also important. However, the consideration of the numerous factors will complicate the processing and computational tasks (e.g., network slicing) in the RAN. Thus, an efficient and smart technique or technology will be required.

\section{Conclusions and Future Directions}
In this article, we described the key drivers for 6G RANs, which are broadly classified into the following use cases: multimedia and entertainment, smart city, smart industry, and beyond-terrestrial communications. Then, we detailed several key performance requirements for 6G RANs, such as flexibility, massive connectivity, and energy efficiency, as well as their corresponding issues. We further emphasized the potential and implementation challenges of network slicing, BS cooperation, NOMA, mMIMO, and energy harvesting in enabling flexible, energy-efficient, and massive radio access for 6G networks.

We observed that the future challenges involve computationally-heavy tasks, such as \emph{(i)} the selection of the best OFDM numerology selection and the optimal placement of logical RAN functions in the flexible RAN slicing problem, \emph{(ii)} beamforming optimization, and \emph{(iii)} power allocation and BS selection for BS cooperation with mMIMO-NOMA. These problems are computationally expensive due to the large dimension of the massive access network (i.e., immerse number of interconnecting nodes and variables). Also, challenges such as the throughput-interference tradeoff problem for energy harvesting is \ally{a multi-objective problem} in nature, which is even more challenging, given the same large-dimensional network scenario. In this case, artificial intelligence (AI) may be a potential solution. Here, we outline several future directions for applying AI approaches to the large-dimensional problems that are difficult to compute:

\ally{\begin{itemize}
\item \emph{AI for network slicing}: AI can be introduced at the cloud and edge servers. The cloud AI can solve problems with higher complexity involving all BSs, such as placement of logical RAN functions at the cloud and OFDM numerology selection. The edge AI can solve more straightforward problems for individual BSs such as resource scheduling among users and cell size expansion. The cloud AI can also interact with the edge AI to jointly instantiate and orchestrate network slice functions and operations. Deep reinforcement learning, which has been shown to have learned the best-slice-allocation policy for network slicing in short time-episodes and achieves better performance than the greedy approaches \cite{drl}, is promising for flexible 6G RANs. 
\item \emph{AI for BS cooperation with mMIMO-NOMA}: By leveraging AI in both the cloud and edge, the cloud AI can determine cooperative BS selection whereas the edge AI can solve the power allocation problem for the users/devices in each mMIMO-NOMA cluster (i.e., a group of users served by a set of cooperating BSs using mMIMO-NOMA).
\item \emph{Edge AI for energy harvesting}: For energy harvesting in a massive access network, distributed AI techniques can be applied at the edge to allow each node to learn and perform the best action (e.g., transmit power level) so as not to degrade the received signal quality of the nodes in proximity.
\end{itemize}}


\ifCLASSOPTIONcaptionsoff
  \newpage
\fi

\bibliographystyle{IEEEtran}
\bibliography{ref}

\end{document}